\documentclass[11pt]{article}

\usepackage{amsfonts,latexsym,eucal,color,graphicx,epsfig,amssymb,amsmath}

\newcommand{\be}{\begin{eqnarray}}
\newcommand{\ee}{\end{eqnarray}}
\newcommand{\beq}{\begin{eqnarray}}
\newcommand{\eeq}{\end{eqnarray}}

\newcommand{\dalm}{\kern1pt\vbox{\hrule height 0.9pt\hbox{\vrule width 0.9pt\hskip 2.5pt\vbox{\vskip 5.5pt}\hskip 3pt\vrule width 0.3pt}\hrule height 0.3pt}\kern1pt}

\usepackage{mathrsfs}

\newcommand{\tr}{\textrm{tr}}
\newcommand{\bh}{\bar{h}}
\newcommand{\bK}{\bar{K}}

\begin{document}

\begin{center}
{\Huge \bf Superselection Sectors of\\
\vspace{.3cm}
Gravitational Subregions}
\vspace{1.5cm}

{\large \bf Joan Camps}

\vspace{.5cm}
{\it Department of Physics and Astronomy, University College London\\
Gower Street, London WC1E 6BT, United Kingdom}\\
\vspace{.5cm}
\verb"j.camps@ucl.ac.uk"

\vspace{2 cm}
{\bf Abstract}

\vspace{.5cm}
\end{center}
Motivated by the problem of defining the entanglement entropy of the graviton, we study the division of the phase space of general relativity across subregions. Our key requirement is demanding that the separation into subregions is imaginary---i.e., that entangling surfaces are not physical. This translates into a certain condition on the symplectic form. We find that gravitational subregions that satisfy this condition are bounded by surfaces of extremal area. We characterise the `centre variables' of the phase space of the graviton in such subsystems, which can be taken to be the conformal class of the induced metric in the boundary, subject to a constraint involving the traceless part of the extrinsic curvature. We argue that this condition works to discard local deformations of the boundary surface to infinitesimally nearby extremal surfaces, that are otherwise available for generic codimension$-2$ extremal surfaces of dimension $\geq 2$.

\newpage
\section{Introduction}
Entanglement entropy in quantum field theory potentially accounts for horizon entropy in gravity \cite{Bombelli:1986rw, Srednicki:1993im}. The Ryu-Takayanagi formula \cite{Ryu:2006bv, Hubeny:2007xt} makes this connection precise in AdS/CFT, where, at large $N$, boundary entanglement entropy is given by areas of bulk extremal surfaces.\footnote{See \cite{Emparan:2006ni} for an example where black hole entropy is fully accounted for by entanglement across the horizon.} It has been argued that, to subleading order in $1/N$, this formula receives quantum corrections \cite{Faulkner:2013ana}:
\beq
S_{\textrm{bdry}}=\frac{A}{4G}+S_{\textrm{bulk}}+\cdots\,,
\label{eq:ST}\eeq
where $S_{\textrm{bulk}}$ denotes the entanglement entropy of bulk fields across the Ryu-Takayanagi surface.

Being the sum of a geometric and a matter term, the right hand side of \eqref{eq:ST} has a natural interpretation as a total bulk entropy. In the context of black holes, such sum is called `generalised entropy' and was introduced by Bekenstein \cite{Bekenstein:1974ax}. One of its virtues is that it should increase during black hole evaporation, thus ensuring that the second law of thermodynamics holds despite the shrinking of the area contribution \cite{Wall:2011hj}.

Other examples of an interplay between geometric and matter contributions to entropic quantities include the quantum focussing conjecture \cite{Bousso:2015mna}, which encompasses covariant entropy bounds, and from which the quantum null energy condition was first conjectured. 

In AdS/CFT, the formula \eqref{eq:ST} can be used to argue that bulk reconstruction from field-theoretic subregions holds in the entanglement wedge \cite{Dong:2016eik}, and that the set of field-theoretic quantum states with a common geometric dual span a quantum error correcting subspace \cite{Almheiri:2014lwa}.

One difficulty with eq.~\eqref{eq:ST} is that $S_{\textrm{bulk}}$ receives contributions from all fields in the bulk---which include, in particular, the graviton. However, as is well known, the entanglement entropy of fields with local symmetries is subtle. The main obstruction is that, while the definition of entanglement entropy assumes a factorising Hilbert space across regions, gauge fields describe extended degrees of freedom that do not naturally factorise. While this problem is already present in electromagnetism, it is exacerbated in general relativity, because there are no local diff-invariant observables. One then needs a prescription for dealing with these non-factorising degrees of freedom.\footnote{See \cite{Harlow:2015lma, Harlow:2018tqv} for discussions of this problem in AdS/CFT wormholes.}

This paper studies subregions in general relativity, and gives a prescription for a notion of factorisation of the graviton across a subregion, and its entanglement with the outside. The main object of analysis is the symplectic form of the graviton. The symplectic form is the basic structure of phase space. In a field theory, this object can be read from the kinetic term. Since the symplectic form is the integral of a local object, our analysis is local. One main conclusion is that, unlike for other fields, with this formalism one can only discuss gravitational subregions bounded by extremal-area surfaces.

The guiding principle that leads to these conclusions is that the subregions of consideration are bounded by `imaginary surfaces'. That is, that the boundaries are not made of any physical substance. This entails that subregions be defined only with the ingredients of the underlying theory. Since gravity is diffeomorphism-invariant, the only separation of subsystems that is allowed has to be defined independently of coordinates. Hence, the local analysis implies that in GR the only allowed entangling surfaces are extremal surfaces.

\section{Symplectic reduction of gauge symmetries}
In this section we review some features of the hamiltonian formalism and gauge symmetries. These ingredients are necessary for the analysis in the rest of this paper.

The phase space of a system is its space of states. One way to think about it is as the space of initial data. More covariantly, phase space is the space of solutions to the equations of motion \cite{Crnkovic:1986ex}.

The main object in phase space is the symplectic form $W$; a non-degenerate, closed, $2-$form
\beq
W(w,v)=I_{v}I_{w}W=-I_{w}I_{v}W\,,\qquad \delta W=0\,,
\eeq
where $I_{v}W$ denotes contraction of a phase space vector field $v$ with the first index of $W$. $\delta$ is the exterior derivative operator in phase space.\footnote{We reserve $i_v$ and $d$ for contraction and exterior derivative of spacetime forms.} For a particle in one dimension $q$, $W=\delta q\wedge \delta p$.

Gauge symmetries are spurious degrees of freedom. In phase space, these non-degrees of freedom appear as null directions $g$ in the would-be symplectic form $W$. Such degenerate $W$ is called a `presymplectic form': it is not a true symplectic form because these can't be degenerate. Often, these directions $g$ are only null in some constraint surface $C$ in phase space:
\beq
\left.I_g W\right|_{C}=0\,.
\eeq

For example, in electromagnetism we often think of phase space as charted by the vector potential and the electric field on a Cauchy slice: $\{A_i(x), E^j(x)\}$. In this space of initial data, the presymplectic form is degenerate on gauge transformations $g_{\varepsilon}=\{\delta A_i(x)=\partial_i\varepsilon(x)\}$ only on the phase space surface satisfying Gauss' law, $C:\{\nabla_iE^i=0\}$.

The true phase space is the quotient of the gauge-redundant space by the orbits of the gauge transformations $g$. This quotient produces the familiar, smaller, space of gauge-inequivalent configurations; and this phase space is equipped with a true, non-degenerate symplectic form $W$, which is inherited from the degenerate presymplectic form in the larger, gauge-redundant space.

This symplectic reduction \cite{Lee:1990nz} proceeds just like ordinary Kaluza-Klein reduction, from the gauge-redundant `phase space' to the true phase space of gauge-invariant states. As in KK, a necessary condition to carry out the reduction is that the gauge directions $g$ are symmetries of the object to be reduced---the presymplectic form. That is, to implement symplectic reduction of $W$ on $C$ over $g$ we need:
\beq
\left.L_g W\right|_{C}=\left.\delta \left(I_g W\right)\right|_{C}=0\,.
\label{eq:SymmW}\eeq
Here $L_g$ denotes the phase space Lie derivative, and we used that $L_g W = \delta I_gW+I_g\delta W$, and that $\delta W=0$.

Condition \eqref{eq:SymmW} is  redundant in principle, but useful in practice. It is redundant because, if one restricts $W$ to $C$ properly, \eqref{eq:SymmW} holds as an identity, as the closedness of $W$ is preserved under restriction and $I_gW$ is identically zero on $C$. The problem in practice, however, is that restricting $W$ to $C$ is not always technically straightforward. In particular, as in the GR case below,  it may not be trivial to decide a priori which directions of phase space are tangent to $C$ and which are not, and hence which legs of $W$ must be kept or discarded under the restriction to $C$.\footnote{This is also the reason to introduce `Dirac brackets' when restricting Poisson brackets (which are defined in terms of $W$) to constraint surfaces $C$. The role of Dirac brackets is to trivialise  discarding gradients away from $C$.}

Eq.~\eqref{eq:SymmW} implies that, on $C$, $I_g W$ is locally exact. One says that $g$ are hamiltonian transformations:
\beq
\left.I_gW\right|_{C}=\delta H_{g}\,.
\eeq

To summarise: in phase space, gauge transformations correspond to null directions $g$ of the presymplectic form $W$. One can reduce to the physical phase space of gauge-inequivalent configurations if $g$ are hamiltonian directions of $W$.

\subsection{Symplectic reduction of gauge transformations on subregions}
Entanglement considerations need, as starting point, the division of a system into subsystems, which may then be entangled. For field theories, natural subsystems are subregions. These are bounded domains of an initial data slice---or, more covariantly, the domains of dependence of these subregions.

Our key physical requirement will be to impose that the subregions are not separated by physical membranes. That is, that this separation of the system into constituents is only an imaginary separation, not a physical one. This is a key defining property of entangling surfaces in field theory.

The demand of the boundary not being physical translates into the condition that the separation of the system into subsystems does not introduce degrees of freedom. If it did introduce degrees of freedom, these would be the degrees of freedom of the separation, which would therefore be a physical membrane; The separation would be made of something---the new degrees of freedom.

As we will review, subregions in gauge theories tend to have boundary degrees of freedom: the gauge transformations with support on the boundary of the subregion.\footnote{See \cite{Donnelly:2016auv} for a rigorous analysis of edge modes in Yang-Mills and GR, and \cite{Speranza:2017gxd} for general diff-invariant theories; and these together with \cite{Harlow:2016vwg} and \cite{Lin:2017uzr, Lin:2018xkj} for constructions recovering the area term in \eqref{eq:ST} from the edge modes.} This is so because boundaries upgrade these gauge transformations from redundant to `large'. In some contexts, as in the Quantum Hall effect, these edge modes are physical \cite{Tong:2016kpv}.

Instead, we will insist that the separations are imaginary; we will seek to avoid edge modes. This involves fixing certain `boundary conditions' on the separating surface. These boundary conditions discard the degrees of freedom that would otherwise become the edge modes.

The role of these boundary conditions is not to fix the state of the fields on the boundary---this would make the boundary physical. Rather, their role is to exclude certain directions from the tangent space of the phase space that we associate to the subregion. These are the directions that would change the state of the boundary, i.e., the boundary degrees of freedom. Discarding them forbids certain motions in phase space, and the state space of a subregion thus decomposes into disconnected superselection sectors labelled by the values of their `boundary conditions' \cite{Jafferis:2015del}. Then, we average over the superselection sectors.

In phase space, this discussion translates to imposing that gauge transformations $g$ with support on the boundary of subregions $\Sigma$, $\left.g\right|_{\partial\Sigma}\neq0$, continue to be non-degrees of freedom of the presymplectic form $W$. That is, that they continue to be null and hamiltonian, so that one can still symplectically reduce over them.

In the next section we will see how these requirements recover known results in electromagnetism, and in the following section we will use them in general relativity.

\section{Electromagnetism}\label{sec:EM}
This section and the next build up on arguments that were first laid down in \cite{Jafferis:2015del}.

The presymplectic form for the Maxwell field is
\beq
W(\delta_1, \delta_2)=\int_{\Sigma}\sqrt{g_{\Sigma}}\,d^{D-1}x \left(\delta_1 A_i\, \delta_2 E^i-
\delta_2 A_i\, \delta_1 E^i\right)\,.
\eeq

Consider this presymplectic form on a  subregion bounded by $\partial \Sigma$. Evaluating $W$ on a gauge direction, $g_{\varepsilon}=\{\delta A_i(x)=\partial_i\varepsilon(x)\}$, gives
\beq
W(g_{\varepsilon}, \delta)=\int_{\Sigma}\sqrt{g_{\Sigma}}\,d^{D-1}x \left(\nabla_i(\varepsilon\, \delta E^i)-\varepsilon\, \nabla_i \delta E^i\right)=\int_{\partial\Sigma}\sqrt{g_{\partial\Sigma}}\, d^{D-2}\sigma\,\varepsilon\, \delta E^\perp\,.
\label{eq:WEM}\eeq
Here we defined $E^\perp\equiv n_i E^i$, with $n_i$ the outward pointing unit normal to $\partial \Sigma$ in $\Sigma$. In the second equality we integrated by parts the first term and used Gauss' law to discard the second term.

We now demand that the separation of $\Sigma$ into the subregions inside and outside $\partial \Sigma$ is imaginary. That is, that $g_\varepsilon$ is a gauge, non-degree of freedom that can be symplectically reduced upon. As argued in the previous subsection, this demands two things on $g_\varepsilon$: that it continues to be null and hamiltonian in the presence of $\partial\Sigma$.

The hamiltonian condition is automatically satisfied in eq.~\eqref{eq:WEM}. Defining
\beq
H_{g_\varepsilon}=\int_{\partial\Sigma} \sqrt{g_{\partial\Sigma}}\,d^{D-2}\sigma\, \varepsilon\,E^\perp\,,
\eeq
we have that\footnote{We consider gauge transformations that are independent of the fields: $\delta \varepsilon=0$.}
\beq
W(g_{\varepsilon}, \delta)=\delta H_{g_\varepsilon}\,,
\eeq
and thus $g_\varepsilon$ is a symmetry direction of the presymplectic form $W$.

For $g_\varepsilon$ to be a null direction of $W$ we need  \eqref{eq:WEM} to vanish. Two natural ways to achieve this are demanding that either $\left.\varepsilon\right|_{\partial\Sigma}=0$, or that $\left.\delta E^{\perp}\right|_{\partial\Sigma}=0$.

These two possibilities are in correspondence with two natural choices in the algebraic discussion of entanglement entropy for gauge fields \cite{Casini:2013rba}: the electric and magnetic centres. The electric centre considers sectors with fixed normal electric field in the boundary. The magnetic centre sector fixes, instead, the tangent components of the vector potential on the boundary: $\left.\delta A_i\right|_{\partial\Sigma}=0$.\footnote{More precisely, the holonomies of the induced $A_i$ on $\partial\Sigma$.} This boundary condition naturally requires that $\left.\varepsilon\right|_{\partial\Sigma}=0$.\footnote{Strictly speaking it only demands that $\varepsilon$ is constant $\varepsilon_0$ on $\partial\Sigma$, but $\varepsilon$ and $\varepsilon-\varepsilon_0$ are the same gauge transformation.}

Notice that there is much more freedom of choice than these two centres: at each point in $\partial \Sigma$ one can choose wether to fix $E^\perp$ or $A_i$ \cite{Casini:2013rba}.


In the algebraic discussion, one chooses the centre and restricts the state to sectors with definite values of the centre operators. For example, one would write the state in $\Sigma$ in the electric centre choice as 
\beq
\rho=\bigoplus_{E^\perp}p_{E^\perp}\rho_{E^\perp}\,,
\label{eq:rhoEl}\eeq
where $\tr\rho_{E^\perp}=1$, and $p_{E^\perp}$ is the probability of the superselection sector with boundary electric field $E^\perp$. A similar decomposition holds for other choices of centres.

The von Neumann entropy of the state \eqref{eq:rhoEl} is
\beq
S_{\textrm{el}}=\sum_{E^\perp} p(E^\perp)S_{E^{\perp}}+H_{E^{\perp}}\,,\qquad H_{E^{\perp}}\equiv-\sum_{E^{\perp}}p_{E^{\perp}}\ln p_{E^{\perp}}\,,
\label{eq:AlgEnt}\eeq
The first term in $S_{\textrm{el}}$ can be interpreted as distillable entropy \cite{Ghosh:2015iwa, Soni:2015yga}, and the second term is the Shannon entropy of the centre variables. Since $E^\perp$ is a functional space, the sums should be interpreted as functional integrals. $p_{E^\perp}$ is really a probability density on this functional space of $E^\perp$s on $\partial\Sigma$.

As discussed in \cite{Donnelly:2014fua, Donnelly:2015hxa}, the Shannon entropy of continuous probability distributions is not necessarily positive. In fact, while the distillable entropy should be independent of the choice of centre variables, the total entropy depends on the choice of centre \cite{Casini:2013rba}.

For abelian lattice gauge theories, the electric centre choice is in correspondence with the `extended Hilbert space' construction of \cite{Buividovich:2008gq, Donnelly:2011hn}, in which one enlarges the Hilbert space with boundary degrees of freedom---edge modes---\cite{Soni:2015yga}. See \cite{Lin:2018bud} for a recent review.

\section{General Relativity}
The symplectic form of general relativity evaluated on a diffeomorphism $\zeta$ is \cite{Jafferis:2015del, Hollands:2012sf}:
\beq
W(g;\delta g, \pounds_\zeta g)=\int_{\partial \Sigma}\delta Q_\zeta(g)-i_\zeta\theta(g;\delta g)\,,
\label{eq:sympFG}\eeq
where we have used the equations of motion: the metrics $g$ and $g+\delta g$ are on-shell.\footnote{In this section $g$ denotes the metric, not a gauge transformation; these are denoted just by their action $\pounds_\zeta g$.}  Just as in electromagnetism, this expression is a boundary term: in the absence of boundaries, diffeomorphisms are trivially null directions of the symplectic form.

In \eqref{eq:sympFG} $Q_{\zeta}$ is the Noether charge density of the diffeomorphism:
\beq
Q_{\zeta}(g)=-\frac{\epsilon_h}{16\pi G}\,\epsilon^{ab}\,\nabla_a\zeta_b\,,
\eeq
where $\epsilon_h$ is the volume $(D-2)-$form of the metric induced on $\partial\Sigma$, and $\epsilon^{ab}$ is its binormal.
$\theta$ is the boundary term that relates the variation of the Lagrangian to the equations of motion upon  integration by parts:\footnote{If $L$ is the Lagrangian $D-$form and $E^{\mu\nu}$ the Einstein equations, $\theta$ is defined by $\delta L=E^{\mu\nu} \delta g_{\mu\nu} +d\theta(g;\delta g)$.}
\beq
i_\zeta\theta(g;\delta g)=\frac{\epsilon_h}{16\pi G}\,v^a(g;\delta g)\, \zeta^b\,\epsilon_{ab}\,,
\eeq
with
\beq
v^a(g;\delta g)\equiv g^{ac}g^{bd}\left(\nabla_d\delta g_{bc}-\nabla_c\delta g_{bd}\right)\,.
\eeq

To evaluate \eqref{eq:sympFG}, we fix the coordinates of the background metric $g$ around $\partial\Sigma$:
\begin{align}
ds^2=&\left(h_{ij}+2K_{ija} x^a+Q_{ijab}\,x^a x^b\right)d\sigma^i d\sigma^j+2 a_{i}\,\epsilon_{ab}\, x^b\, dx^a\,d\sigma^i \nonumber\\
&-\frac{4}{3}R_{iabc}\,x^a x^b dx^c d\sigma^i+\left(\eta_{bd}-\frac{1}{3}R_{abcd}\,x^a x^c\right)dx^b dx^d+O(x^3)\,.
\label{eq:AdaptedCoords}
\end{align}
In these coordinates the separating surface $\partial\Sigma$ lies at $x^a=0$, $a=0,1$,\footnote{Until now, $a,b$ were generalised indices. From now on, they relate to this choice of coordinates. Likewise, $i, j$ indices spanned $\Sigma$ in sec.~\ref{sec:EM}; now they span $\partial\Sigma$.} and $\eta_{ab}=\textrm{diag}(-1,1)$. $x^a$ are normal coordinates away from $\partial\Sigma$, and eq.~\eqref{eq:AdaptedCoords} makes explicit only the dependence on $x^a$---all objects depend implicitly on $\sigma^i$, the coordinates on  $\partial\Sigma$. $h_{ij}$ is the intrinsic metric of $\partial\Sigma$ and $K_{ija}$ its extrinsic curvature. The binormal of $\partial\Sigma$ is $\epsilon_{ab}\,dx^a dx^b=dx^0\wedge dx^1$.

It will be convenient to divide the intrinsic metric and extrinsic curvature of $\partial\Sigma$ into their trace and traceless parts:
\beq
h_{ij}\equiv e^{2\Omega}\bh_{ij}\,,\qquad \det \bh_{ij}=1\,,
\label{eq:defhb}\eeq
so that the volume form in $\partial\Sigma$ is $\epsilon_h=e^{(D-2)\Omega}d\sigma^1\wedge\cdots
\wedge d\sigma^{D-2}$, and
\beq
K_{ija}\equiv\frac{1}{D-2}h_{ij}K_a+e^{2\Omega}\bK_{ija}\,,\qquad h^{ij}\bK_{ija}=0\,.
\label{eq:defKb}\eeq
Notice that the definitions \eqref{eq:defhb}, \eqref{eq:defKb} imply that variations satisfy
\beq
h^{ij}\delta \bh_{ij}=0\,,\qquad \bh^{ij}\,\delta\bK_{ija}=\bK^{ija}\,\delta \bh_{ij}\,.
\eeq

With these definitions, eq.~\eqref{eq:sympFG} evaluates to:\footnote{Restoring to Euclidean signature, this corrects expression (A.28) in \cite{Jafferis:2015del}.}
\begin{align}
W(g;\delta g, \pounds_\zeta g)=
\frac{1}{8\pi G}\int_{\partial \Sigma}\bigg[-\zeta^i\delta (a_i\,\epsilon_h)+\zeta^\tau\delta(\epsilon_h)+
\frac{1}{D-2}\zeta^b\epsilon^a{}_{b}\,\delta(K_a\,\epsilon_h)&\cr
+\left(\frac{D-3}{D-2}\delta K_a\,+\frac{1}{2}\bar{K}^{ij}{}_{a}\,\delta\bar{h}_{ij}\right) \zeta^b \epsilon^a{}_{b}\,\epsilon_h&\bigg]\,,
\label{eq:sympN}
\end{align}
where $\partial_\tau\equiv x^1\partial_{x^0}+x^0\partial_{x^1}$.

There are three types of diffeomorphisms contributing to $W(g;\delta g,\pounds_\zeta g)$: surface diffeomorphisms $\zeta^i$, boosts $\zeta^\tau$, and translations $\zeta^a$. The first line of \eqref{eq:sympN} shows hamiltonians for each of these types of generators; the second line is an apparent obstruction to translations $\zeta^a$ being hamiltonian.

Now we impose that the entangling surface is imaginary; this demands that diffeos continue to be null, and hamiltonian, in the presence of $\partial\Sigma$.

Consider first the surface diffeomorphisms $\zeta^i$. There are two natural ways to drop them from $W$; One is to not let $a_i\,\epsilon_h$ fluctuate. This is analogous to the electric boundary conditions for electromagnetism. The other way is analogous to the magnetic choice:  fix instead the conformal class of the induced metric on $\partial\Sigma$, $\delta\bh_{ij}=0$\,---\,$\bh_{ij}$ would transform under a diffeomorphism on the surface, so fixing $\bh_{ij}$ generically sets $\zeta^i=0$.\footnote{except for conformal Killing vectors, which are not generic, and are at best a finite dimensional subspace of $\zeta^i(\sigma)$ in $D>4$. In $D=4$, they make the infinite-dimensional conformal group of $2-$dimensional surfaces. 
These CKVs, however, have one-dimensional dependence, and so still are a zero-measure subset of $\zeta^i(\sigma^1, \sigma^2)$.} In this paper we focus on the option that fixes $\bh_{ij}$.

Let us now look at surface translations $\zeta^a$. Their effect is to slightly displace $\partial\Sigma$. It is then natural that they will drop from $W$ if the location of $\partial\Sigma$ is sufficiently rigidly specified, so that no such deformations are active. Indeed, the translation term drops from the first line of \eqref{eq:sympN} if we do not let $K_a\,\epsilon_h$ fluctuate. One natural way to achieve this is demanding that $\partial\Sigma$ has extremal area, $K_a=0$, and that fluctuations respect this extremality: $\delta K_a=0$.

$\delta K_a=0$ also gets rid of the first term in the second line of \eqref{eq:sympN}. Recall that this second line is a potential obstruction to $\zeta^a$ being hamiltonian, and making it zero ensures that we are properly restricting $W$ to a constraint surface in phase space---in this case, a certain $K_a=0$ locus. To achieve this restriction we also need to demand the vanishing of the last term in eq.~\eqref{eq:sympN}:
\beq
\bK^{ij}{}_{a}\,\delta \bh_{ij}=0\,.
\label{eq:bKdh}
\eeq
This suggests that the extremality condition $K_a=0$, $\delta K_a=0$, does not fully specify the location of the entangling surface $\partial \Sigma$. Indeed, we argue below that, generically, codimension$-2$ extremal surfaces can be infinitesimally locally deformed. As we will see, condition \eqref{eq:bKdh} then discards this freedom.

Observe that the hamiltonian for homogeneous boosts $\zeta^\tau$ is 
\beq
H_{\zeta^\tau}=\frac{\left.\textrm{Area}\right|_{\partial\Sigma}}{8\pi G}\,\zeta^\tau\,.
\eeq
We could drop these by demanding that the area of $\partial\Sigma$ is fixed, but we will not do that in this paper. One reason not to do it is that $\partial_\tau=0$ on $\partial\Sigma$, so $\zeta^\tau$ does not act on the boundary. Another reason is by comparison to the codimension$-1$ problem, where the trace of the extrinsic curvature is canonically conjugate to the area, so fixing both would be overconstraining, and we have already fixed $\delta K_a=0$.

The fact that boosts are physical symmetries in this formalism may be related to the physicality of modular flow---which, for ordinary quantum fields, acts as a boost near the entangling surface.

In summary, extremal surfaces $\partial\Sigma$ make good entangling surfaces in general relativity. The superselection sectors on which phase space splits can be taken to be labelled by the induced conformal metric $\bh_{ij}$, subject to the constraint \eqref{eq:bKdh}.

In this choice of centre, the state of the graviton in $\Sigma$ decomposes into:
\beq
\rho_{\Sigma}=\bigoplus_{\{\delta \bh_{ij}|\bK^{ija}\,\delta\bh_{ij}=0\}} p_{\delta\bh_{ij}}\,\rho_{\delta\bh_{ij}}\,,
\label{eq:rhoGrav}\eeq
where $\rho_{\delta\bh_{ij}}$ are normalised states of the degrees of freedom in $\Sigma$ subject to making $\partial\Sigma$ a minimal surface, with induced conformal metric $\bh_{ij}+\delta \bh_{ij}$.\footnote{Since the analysis is perturbative, the direct sum is over fluctuations $\delta\bh_{ij}$. $\bh_{ij}$ is the metric induced on $\partial\Sigma$ in the classical background.} The entropy of \eqref{eq:rhoGrav} is just like in \eqref{eq:AlgEnt}.\footnote{This requires a measure in the space of constrained $\bar{h}_{ij}$s. Defining this type of measures is subtle \cite{Lin:2018xkj}.}

\subsection{Generic local deformability of codimension$-2$ extremal surfaces}
We now sketch an argument why codimension$-2$ extremal surfaces are generically infinitesimally deformable.  Such a deformation of $\partial\Sigma$ from $x^a=0$ to a nearby $x^a=\zeta^a(\sigma)$ is called a Jacobi field, and satisfies the equation $\delta_{\zeta}K^a=0$:
\beq
\delta_{\zeta} K^a=-D_i\left(D^{i}\zeta^a\right)+Q^\prime{}_b{}^a\zeta^b=0\,,
\label{eq:deltaK}\eeq
where
\beq
Q^{\prime}{}_{ij}{}^{ab}\equiv Q_{ij}{}^{ab}-2K_{(i}{}^{ka}K_{j)k}{}^{b}+a_{i}\, a_{j}\,\eta^{ab}\,,
\label{eq:Qprime}
\eeq
and $Q^\prime{}_b{}^a\equiv h^{ij}Q^{\prime}{}_{ijb}{}^a$. The $D_i$ derivative is
\beq
D_i\zeta^a\equiv \nabla_i\zeta^a-a_i\,\epsilon^a{}_b\,\zeta^b\,,
\label{eq:covD}\eeq
where $\nabla_i$ is the standard covariant derivative in $\partial\Sigma$ compatible with its induced metric $h_{ij}$, and is transparent to $a-$indices. The $a_i$ object of \eqref{eq:AdaptedCoords} acts as a connection along $\partial\Sigma$ for boosts of the normal directions.

Eq.~\eqref{eq:deltaK} reduces to a Laplace equation for $\zeta^a$ when $\partial\Sigma$ is a flat surface embedded in Minkowski space, as one expects. Inspection of the adapted coordinates \eqref{eq:AdaptedCoords} also makes the appearance of $Q_{ijab}\,\zeta^b$  plausible in the variation of $K_{ija}$ under the deformation generated by $\zeta^a$.

Infinitesimal local deformability of extremal surfaces is a statement about existence of solutions to eq.~\eqref{eq:deltaK} centred on arbitrary points $\sigma^i=\bar{\sigma}^i$. If such solutions decay fast enough so that they can be considered localised around $\bar{\sigma}^i$, they can be called local deformations. In AdS/CFT, these would be infinitesimal deformations of the HRT surface that do not change the boundary anchoring surface.

We will argue for the existence of localised solutions to \eqref{eq:deltaK}. This involves analysing \eqref{eq:Qprime}, which can be simplified with Gauss-Codacci-type of equations, that relate the $Q_{ijab}$ and $a_i$ objects to the background Riemann tensor of \eqref{eq:AdaptedCoords} on $\partial\Sigma$:
\begin{align}
R_{ij}{}^{kl} &= {\cal R}_{ij}{}^{kl} - 2 K_{i}{}^{[k}{}_{c}\, K^{l]}{}_{j}{}^{c} \cr
R_{ij}{}^{ab} &= F_{ij}\,\epsilon^{ab} - 2 K_{[i}{}^{ka}\, K_{j]k}{}^{b} \cr
R_{(i}{}^a{}_{j)}{}^{b} &= - Q^\prime{}_{ij}{}^{ab}-K_{(i}{}^{ka}\, K_{j)k}{}^{b}\,,
\label{eq:GC}\end{align}
where ${\cal R}_{ijkl}$ is the Riemann tensor of $h_{ij}$, the intrinsic metric on $\partial \Sigma$, and $F_{ij}=2\partial_{[i}a_{j]}$ is the curvature of the abelian connection on normal boosts, $a_i$.

Using \eqref{eq:GC} we can rewrite the trace of \eqref{eq:Qprime} as
\beq
Q^\prime{}_{ab}\equiv\frac{1}{2}\left(-h^{ij}R_{ij}+ {\cal R} 
+(\bar{K}^2){}^c{}_c\right)\eta_{ab}-R_{\{ab\}}-(\bar{K}^2){}_{ab}
\label{eq:fullQprime}\eeq
where we used that $\partial\Sigma$ is a minimal surface, $K_a=0$, and defined $R_{\{ab\}}\equiv R_{ab}-\frac{1}{2}\,R_{cd}\,\eta^{cd}\,\eta_{ab}$, and $(\bar{K}^2){}^a{}_{b}\equiv \bar{K}^{ika}\, \bar{K}_{ikb}$.\footnote{The Raychaudhuri equation for congruences of null geodesics emanating from an extremal surface can be obtained from \eqref{eq:fullQprime} by specialising $ab$ to a null direction, say $v$, and recognising that $Q^{\prime}_{vv}=\dot{\theta}_{(v)}$, and that $\bar{K}_{ijv}=\sigma_{(v)ij}$. We then have the familiar $\dot{\theta}_{(v)}=-R_{vv}-(\sigma_{(v)})^2$. The usual $-\frac{1}{D-2}(\theta_{(v)})^2$ term is missing because of extremality: $\theta_{(v)}=0$.}

Making the approximation that the induced metric is conformally flat $\bh_{ij}=\delta_{ij}$, which in $D=4$ is not an approximation,\footnote{Two-dimensional surfaces are conformally flat.} we can conformally transform $h_{ij}$ to $\delta_{ij}$ in \eqref{eq:deltaK} and write the kinetic term as in flat space:
\beq
D_i \left(D^i\zeta^a\right)-\frac{D-4}{4(D-3)}{\cal R}\,\zeta^a=e^{-\frac{D}{2}\Omega}\delta^{ij}\left(\partial_i-a_i\right)
\left(\partial_j-a_j\right)\tilde{\zeta}^a
\eeq
where $\tilde{\zeta}^a=e^{\frac{D-4}{2}\Omega}\zeta^a$.

In this way \eqref{eq:deltaK} becomes a flat space equation on $\partial\Sigma$
\beq
-\delta^{ij}\left(\partial_i-a_i\right)
\left(\partial_j-a_j\right)\tilde{\zeta}^a+V^a{}_b\,\tilde{\zeta}^b=0\,,
\label{eq:Sch}\eeq
with:
\beq
V^a{}_b=e^{2\Omega}\left[\frac{1}{2}\left(-h^{ij}R_{ij}+ \frac{1}{2}\frac{D-2}{D-3}{\cal R} 
+(\bar{K}^2){}^c{}_c\right)\delta^{a}{}_{b}-\eta^{ac}R_{\{cb\}}-(\bar{K}^2){}^a{}_{b}\right]\,.
\eeq

Ignoring $a_i$, eq.~\eqref{eq:Sch} is a zero-energy Schroedinger-type equation for a vector quantity $\tilde{\zeta}^a$, with potential $V^a{}_b$.\footnote{Keeping $a_i$, eq.~\eqref{eq:Sch} is a zero-energy Schroedinger-type equation for a charged particle in a magnetic field $F_{ij}$, under a complex potential that also acts on the complex conjugate of the `wave function' $\zeta^0+i\,\zeta^1$ \cite{Camps:ToAppear}.} Localised zero-energy solutions are only possible if the potential attains negative values.

To analyse this further, let us consider a region where the background is approximately flat, so the Ricci terms vanish and ${\cal R}=-(\bar{K}^2){}^c{}_c$, giving
\beq
{}^{\textrm{flat}}V^a{}_b(\sigma)=\frac{e^{2\Omega}}{4}\left[\frac{D-4}{D-3}(\bar{K}^2){}^c{}_c\,\delta^a{}_b
-4(\bar{K}^2){}^a{}_b\right]\,,
\label{eq:Vab}\eeq
and
\beq
{}^{\textrm{flat}}F_{ij}=-\bK_{i}{}^{ka}\,\bK_{jk}{}^{b}\,\epsilon_{ab}\,.
\eeq

$(\bar{K}^{2}){}^{a}{}_b$ has a two eigenvalues, $(\bar{\kappa}_{\hat{1}})^2$ and $-(\bar{\kappa}_{\hat{0}})^2$, for a spacelike and timelike eigenvectors. Since $(\bar{K}^{2}){}^{a}{}_b$ is a square, the spacelike eigenvalue is non-negative, and the timelike one is non-positive.

The spacelike eigenvalue of $V^a{}_b$ then is:
\beq
V_{\hat{1}}=-\frac{e^{2\Omega}}{4}\left[\frac{D-4}{D-3}(\bar{\kappa}_{\hat{0}})^2+\frac{3D-8}{D-3}(\bar{\kappa}_{\hat{1}})^2\right]\,,
\eeq
and is negative in $D=4$ if $\bar{\kappa}_{\hat{1}}\neq 0$, which is generic, and more generally for $D>4$ if $\bK_{ija}\neq 0$.

Approximating such eigenvalue by a constant, $V_{\hat{1}}\approx-\bar{\kappa}^2$, eq.~\eqref{eq:Sch} in the direction $\hat{1}$ becomes:
\beq
-\partial^i\partial_i\,\tilde{\zeta}^{\hat{1}}-\bar{\kappa}^2\,\tilde{\zeta}^{\hat{1}}=0\,,
\label{eq:Schm}
\eeq
where we are ignoring $F_{ij}$, and thus $a_i$. Eq.~\eqref{eq:Schm} does admit spherical wave solutions, decaying away from a centre. For example, in $D=4$, we have
\beq
\tilde{\zeta}^{\hat{1}}= J_0(\bar{\kappa}\,\sigma)\,,\qquad \sigma\equiv\sqrt{\delta_{ij}(\sigma^i-\bar{\sigma}^i)
(\sigma^j-\bar{\sigma}^j)}\,,
\label{eq:zetaSol}\eeq
with $\bar{\sigma}^i$ integration constants, and $J_0(\bar{\kappa}\,\sigma)$ Bessel functions, decaying as $\left(\bar{\kappa}\,\sigma\right)^{-1/2}$. \eqref{eq:zetaSol} are then localised deformations to infinitesimally nearby extremal codimension$-2$ surfaces.

To justify neglecting $F_{ij}$ instead of, e.g., approximating it by a constant, recall that a constant magnetic field on a plane acts like a harmonic potential $\propto\sigma^2$, giving rise to Landau levels. $F_{ij}$ should be then kept at the same order as the one that keeps harmonic corrections to $V^a{}_b$ $\propto\sigma^2$, that is, subleading order.

The key ingredient enabling local deformations of $\partial \Sigma$ is a non-zero traceless extrinsic curvature $\bK_{ija}\neq 0$. We will now see that, as announced, condition \eqref{eq:bKdh}, which is active also only when $\bK_{ija}\neq 0$, works to discard these deformations.

An infinitesimal deformation $\zeta^a$ of $\partial \Sigma$ would change the induced metric by
\beq
\delta_{\zeta} h_{ij}=2 K_{ija}\,\zeta^a\,.
\eeq
Substituting this into \eqref{eq:bKdh} we have
\beq
\bK^{ij}{}_a\,\delta_\zeta \bh_{ij}=2\bK^{ij}{}_a\,\bK_{ijb}\,\zeta^b=0\,,
\eeq
which, if $\bK_{ija}\neq0$, can only be satisfied by $\zeta^a=0$. \eqref{eq:bKdh} thus discards the deformations \eqref{eq:zetaSol}.

In fact, eq.~\eqref{eq:bKdh} discards all translations $\zeta^a\neq0$, not only the ones reaching nearby extremal surfaces in the original geometry \eqref{eq:zetaSol}. We will now interpret why all such translations need to be discarded. The reason is that they do reach a nearby extremal surface, albeit possibly in a perturbed geometry. This effect is second order in gravitational perturbations.

Recall that in eq.~\eqref{eq:rhoGrav} the state of a gravitational subregion is given as a mixture of sectors. Each  sector is constituted by a family of background geometries with an entangling extremal surface of interest, at $x^a=0$, with fixed induced conformal metric $\bh_{ij}$, but with the remaining geometric data varying: these are the background Weyl tensor,\footnote{the part of the curvature not fixed by the Einstein equations} as well as the traceless extrinsic curvature and area density of the entangling surface.

The deformation equation \eqref{eq:Sch} has a different potential $V^a{}_b$ within each such geometry and, as a result, its solutions $\tilde{\zeta}^a$ will have varying direction and support depending on this varying data. We expect that, through these variations, translations to nearby entangling extremal surfaces $\zeta^a$ span all directions in the normal plane $ab$ and all supports on $\partial\Sigma$ within a given $\bh_{ij}$ sector.

Such translations are discarded by $\bK^{ija}\,\delta\bh_{ij}=0$ even when the traceless extrinsic curvature has changed to $\bK_{ija}+\delta\bK_{ija}$. To see this, decompose $\delta \bK_{ija}$ into a sum of terms with and without zero product and trace with $\bK_{ija}$ in the $ij$ indices:
\beq
\delta\bK_{ija}\equiv \bK_{ijb}\,\delta_1S^b{}_a+\delta_2 \bK_{ija}\,, \qquad \bK^{ija}\,\delta_2\bK_{ijb}=0\,.
\eeq
To leading order in $\delta \bK_{ija}$, $\delta_2 \bK_{ija}$ decouples from the discussion: On one hand, the variation of the potential in eq.~\eqref{eq:Sch} is
\beq
\delta V^a{}_b\propto\bK^{ija}\,\bK_{ijc}\,\delta_1S^c{}_b +O((\delta_2)^2)\,;
\label{eq:deltaV}
\eeq
and, on the other hand, the variation in eq.~\eqref{eq:bKdh} induced by a translation $\zeta^a$ is
\beq
\bK^{ija}\,\delta \bh_{ij}=2\bK^{ija}
\left(\bK_{ijc}(\delta^c{}_b+\delta_1 S^c{}_b)+\delta_2\bK_{ijb}\right)\zeta^b
=2(\bK^2)^a{}_c\left(\delta^c{}_b+\delta_1 S^c{}_b\right)\zeta^b\,.
\label{eq:dbKdbh}\eeq
$\delta_2\bK_{ija}$ drops both from \eqref{eq:deltaV} and \eqref{eq:dbKdbh}. Because of this, to $O(\delta \bK_{ija})$, eq.~\eqref{eq:bKdh} discards all relevant nearby entangling surfaces, that effectively live within the family of geometries with traceless extrinsic curvatures $\bK_{ijc}(\delta^c{}_b+\delta_1 S^c{}_b)$.

Condition \eqref{eq:bKdh} is, then, generically about second variations: deformations of the entangling surface in a perturbed geometry. That this should be the case can be justified by re-examining its origin in eq.~\eqref{eq:sympN}. We demanded $\bK^{ij}{}_a\,\delta\bh_{ij}=0$ so that the variations generated by $\zeta^a$ were Hamiltonian. $\bK^{ij}{}_a\,\delta\bh_{ij}$ is an obstruction to finding a Hamiltonian only if $\bK_{ija}$ is allowed to depend on phase space, so that $\bK^{ij}{}_a\,\delta\bh_{ij}\neq\delta\left(\bK^{ij}{}_a\,\bh_{ij}\right)$, and this obstruction is so only if $\bK_{ija}$ can have a non-trivial variation in the presence of $\delta\bh_{ij}$. This is a condition of variations in the presence of variations, and hence a second order condition.

These arguments are perturbative, but it is only when the states in \eqref{eq:rhoGrav} spread over small such perturbations of one geometry that we can talk about there being a `background geometry' on which we can consider the entanglement of the graviton across a certain surface.

We have in effect argued that all deformations $x^a=\zeta^a$ of the original extremal entangling surface, at $x^a=0$, can be nearby entangling surfaces, in the original geometry or in a fluctuated one. The role of eq.~\eqref{eq:bKdh} is to discard, from the sum over centres $\bh_{ij}$, entangling surfaces that lie in background geometries that have already been accounted for in the sector with $\delta\bh_{ij}=0$. Indeed, eq.~\eqref{eq:rhoGrav} should describe the state of the graviton across one entangling surface, at $x^a=0$, not across multiple entangling surfaces within each ambient background geometry.

We close this subsection with three comments.

First, having solutions to the Jacobi equation does not imply existence of finite deformations to nearby entangling surfaces.  Any result following from the Jacobi equation is just a first order result.

Second, the sketch of this subsection does not imply anything on geodesics---or dimension$-1$ extremal curves---as for these the enabling term $\bK_{ija}$ is identically zero. But we expect the essence of the argument to hold generically for dimension $>1$ extremal surfaces.

Third, nothing significant in this subsection changes in Euclidean signature; only $\eta_{ab}\rightarrow\delta_{ab}$,  the sign of the $a_i\, a_j$ term in \eqref{eq:Qprime}, and the sign of $H_{\zeta^\tau}$. ${}^{\textrm{flat}}V^a{}_b$ also has generically at least one negative eigenvalue, as can be seen by taking the trace, ${}^{\textrm{flat}}V^a{}_a=-
\frac{e^{2\Omega}}{2}\frac{D-2}{D-3}(\bar{K}^2){}^a{}_a$, which is negative when $\bK_{ija}\neq 0$ because it is minus a sum of squares.

\section{Discussion}
In this paper we have studied entangling surfaces for the gravitational field. Our key physical requirement has been to demand that these surfaces are not physical, i.e., that they do not have degrees of freedom. This translates to the mathematical condition that diffeomorphisms with support on the entangling surface are null and hamiltonian directions of the symplectic form.

When the only field available is the graviton, the analysis outputs extremal surfaces $K_a=0$ as the only type of entangling surfaces for which these properties hold.\footnote{See \cite{Kirklin:2018gcl} for a similar conclusion from an argument with Euclidean path integrals.} We have only examined in some detail one choice of centre variables on which the phase space splits under the separation into subregions; in this choice, the centre variables are the conformal class of the induced metric on the entangling surface, $\delta \bh_{ij}$, subject to the constraint $\bK^{ij}{}_{a}\,\delta\bh_{ij}=0$. We sketched an argument that this constraint works to discard fluctuations to nearby, extremal, entangling surfaces.

Notice that the constraint $\bK^{ij}{}_{a}\,\delta\bh_{ij}=0$ trivialises on bifurcation surfaces of Killing horizons, such as the $r=2M$ surface in Schwarzschild, where $K_{ija}=0$. These surfaces are expected to be good graviton entangling surfaces. But the constraint will be active, for example, in the generic Hubeny-Rangamani-Takayanagi setups \cite{Hubeny:2007xt}.

The analysis of this paper suggests that the entanglement of the graviton may be definable only across surfaces of extremal area $K_a=0$, and this raises questions about some proposed algorithms to compute gravitational entropy to all orders in $G$. According to the prescription in \cite{Engelhardt:2014gca} (see also \cite{Dong:2017xht}), one should evaluate $S_{\textrm{tot}}=A/4G+S_{\textrm{bulk}}$ on surfaces that minimise it, $\delta S_\textrm{tot}=0$. But if $S_{\textrm{bulk}}$ is only defined across extremal surfaces, $\delta$ in $\delta S_\textrm{tot}$ can not explore all surfaces nearby\footnote{I thank Aron Wall for pointing this out.}---at best it would explore extremal surfaces nearby.

Similarly, the condition that the entangling surface be extremal seems to rule out general sections of time-dependent black hole horizons, which make the setup of the generalised second law of black hole thermodynamics. If one is to argue that $S_{\textrm{tot}}$ increases at all steps of black hole evaporation, it ought to be possible to define entanglement of the graviton across non-minimal surfaces.

Continuing on the subject of black holes, it would be very interesting to develop a real-time picture of the quantum corrections to black hole entropy (see \cite{Sen:2007qy} for a review). By importing the AdS/CFT derivation of \cite{Faulkner:2013ana}, one expects that such quantum corrections are given by the entanglement of the graviton $S_{\textrm{bulk}}$ \cite{Solodukhin:2011gn}. But, as briefly recalled in section \ref{sec:EM}, the entanglement entropy of gauge fields---including the so-called `universal' logarithmic corrections---depends on how one chooses the centre \cite{Casini:2013rba}.\footnote{See \cite{Huang:2014pfa, Casini:2015dsg, Huerta:2018xvl} for examples of calculations of entanglement entropy in electromagnetism whose logarithmic divergences do not agree with the ones of the electric centre \cite{Donnelly:2014fua, Donnelly:2015hxa}. The form of the latter is dictated by the central charges. However, see \cite{Donnelly:2016mlc} for an agreement between electric and magnetic calculations.} It is important to understand if this ambiguity continues to apply for the graviton, and if it does, whether it gives rise to any ambiguity in the logarithmic corrections to black hole entropy. While these caveats about logarithms would apply to finite temperature black holes, e.g.~as in Schwarzschild \cite{Sen:2012dw}, they may not be relevant for extremal black holes, for which the entanglement-across-the-horizon picture may break down, as the horizon is infinitely far away.

The fact that we charted the boundary values of fluctuations by the change in the trace of the extrinsic curvature $\delta K_a$ and the change in the conformal class of the induced metric $\bh_{ij}$ is reminiscent to analogous boundary conditions for codimension$-1$ surfaces in Euclidean quantum gravity \cite{Witten:2018lgb}---where it was argued that the more familiar Dirichlet boundary conditions $\delta h_{ij}=0$ fail to make the graviton fluctuation operator elliptic. We note that both there and here the guiding principle in choosing boundary conditions is to discard diffeomorphisms with support on the boundary as degrees of freedom. It would be very interesting to flesh out the extent of this apparent connection more precisely.

\subsection*{Acknowledgments}
It is a pleasure to thank William Donnelly, Aitor Lewkowycz and Antony Speranza for many crucial discussions, feedback, and collaboration on related projects. I also thank Roberto Emparan, Ted Jacobson, and Jonathan Oppenheim for discussions, and Christos Mantoulidis and Aron Wall for correspondence. Parts of this work were done during visits to the Mainz Institute for Theoretical Physics and to the Galileo Galilei Institute, Florence, whose hospitality I acknowledge gratefully. Work supported by the `It from Qubit' Simons collaboration.


\begin{thebibliography}{99}
\bibitem{Bombelli:1986rw}
  L.~Bombelli, R.~K.~Koul, J.~Lee and R.~D.~Sorkin,
  ``A Quantum Source of Entropy for Black Holes,''
  Phys.\ Rev.\ D {\bf 34} (1986) 373.
  doi:10.1103/PhysRevD.34.373

\bibitem{Srednicki:1993im}
  M.~Srednicki,
  ``Entropy and area,''
  Phys.\ Rev.\ Lett.\  {\bf 71} (1993) 666
  doi:10.1103/PhysRevLett.71.666
  [hep-th/9303048].

\bibitem{Ryu:2006bv}
  S.~Ryu and T.~Takayanagi,
  ``Holographic derivation of entanglement entropy from AdS/CFT,''
  Phys.\ Rev.\ Lett.\  {\bf 96} (2006) 181602
  doi:10.1103/PhysRevLett.96.181602
  [hep-th/0603001].

\bibitem{Hubeny:2007xt}
  V.~E.~Hubeny, M.~Rangamani and T.~Takayanagi,
  ``A Covariant holographic entanglement entropy proposal,''
  JHEP {\bf 0707} (2007) 062
  doi:10.1088/1126-6708/2007/07/062
  [arXiv:0705.0016 [hep-th]].

\bibitem{Emparan:2006ni}
  R.~Emparan,
  ``Black hole entropy as entanglement entropy: A Holographic derivation,''
  JHEP {\bf 0606} (2006) 012
  doi:10.1088/1126-6708/2006/06/012
  [hep-th/0603081].
  
\bibitem{Faulkner:2013ana}
  T.~Faulkner, A.~Lewkowycz and J.~Maldacena,
  ``Quantum corrections to holographic entanglement entropy,''
  JHEP {\bf 1311} (2013) 074
  doi:10.1007/JHEP11(2013)074
  [arXiv:1307.2892 [hep-th]].

\bibitem{Bekenstein:1974ax}
  J.~D.~Bekenstein,
  ``Generalized second law of thermodynamics in black hole physics,''
  Phys.\ Rev.\ D {\bf 9} (1974) 3292.
  doi:10.1103/PhysRevD.9.3292

\bibitem{Wall:2011hj}
  A.~C.~Wall,
  ``A proof of the generalized second law for rapidly changing fields and arbitrary horizon slices,''
  Phys.\ Rev.\ D {\bf 85} (2012) 104049
   Erratum: [Phys.\ Rev.\ D {\bf 87} (2013) no.6,  069904]
  doi:10.1103/PhysRevD.87.069904, 10.1103/PhysRevD.85.104049
  [arXiv:1105.3445 [gr-qc]].

\bibitem{Bousso:2015mna}
  R.~Bousso, Z.~Fisher, S.~Leichenauer and A.~C.~Wall,
  ``Quantum focusing conjecture,''
  Phys.\ Rev.\ D {\bf 93} (2016) no.6,  064044
  doi:10.1103/PhysRevD.93.064044
  [arXiv:1506.02669 [hep-th]].

\bibitem{Dong:2016eik}
  X.~Dong, D.~Harlow and A.~C.~Wall,
  ``Reconstruction of Bulk Operators within the Entanglement Wedge in Gauge-Gravity Duality,''
  Phys.\ Rev.\ Lett.\  {\bf 117} (2016) no.2,  021601
  doi:10.1103/PhysRevLett.117.021601
  [arXiv:1601.05416 [hep-th]].
  
\bibitem{Almheiri:2014lwa}
  A.~Almheiri, X.~Dong and D.~Harlow,
  ``Bulk Locality and Quantum Error Correction in AdS/CFT,''
  JHEP {\bf 1504} (2015) 163
  doi:10.1007/JHEP04(2015)163
  [arXiv:1411.7041 [hep-th]].

\bibitem{Harlow:2015lma}
  D.~Harlow,
  ``Wormholes, Emergent Gauge Fields, and the Weak Gravity Conjecture,''
  JHEP {\bf 1601} (2016) 122
  doi:10.1007/JHEP01(2016)122
  [arXiv:1510.07911 [hep-th]].
  
\bibitem{Harlow:2018tqv}
  D.~Harlow and D.~Jafferis,
  ``The Factorization Problem in Jackiw-Teitelboim Gravity,''
  arXiv:1804.01081 [hep-th].
  
\bibitem{Crnkovic:1986ex}
  C.~Crnkovic and E.~Witten,
  ``Covariant Description Of Canonical Formalism In Geometrical Theories,''
  In *Hawking, S.W. (ed.), Israel, W. (ed.): Three hundred years of gravitation*, 676-684 and Preprint - Crnkovic, C. (86,rec.Dec.) 13 p
  
\bibitem{Lee:1990nz}
  J.~Lee and R.~M.~Wald,
  ``Local symmetries and constraints,''
  J.\ Math.\ Phys.\  {\bf 31} (1990) 725.
  doi:10.1063/1.528801

\bibitem{Donnelly:2016auv}
  W.~Donnelly and L.~Freidel,
  ``Local subsystems in gauge theory and gravity,''
  JHEP {\bf 1609} (2016) 102
  doi:10.1007/JHEP09(2016)102
  [arXiv:1601.04744 [hep-th]].
  
\bibitem{Speranza:2017gxd}
  A.~J.~Speranza,
  ``Local phase space and edge modes for diffeomorphism-invariant theories,''
  JHEP {\bf 1802} (2018) 021
  doi:10.1007/JHEP02(2018)021
  [arXiv:1706.05061 [hep-th]].

\bibitem{Harlow:2016vwg}
  D.~Harlow,
  ``The RyuÐTakayanagi Formula from Quantum Error Correction,''
  Commun.\ Math.\ Phys.\  {\bf 354} (2017) no.3,  865
  doi:10.1007/s00220-017-2904-z
  [arXiv:1607.03901 [hep-th]].

\bibitem{Lin:2017uzr}
  J.~Lin,
  ``Ryu-Takayanagi Area as an Entanglement Edge Term,''
  arXiv:1704.07763 [hep-th].

\bibitem{Lin:2018xkj}
  J.~Lin,
  ``Entanglement entropy in Jackiw-Teitelboim Gravity,''
  arXiv:1807.06575 [hep-th].

\bibitem{Tong:2016kpv}
  D.~Tong,
  ``Lectures on the Quantum Hall Effect,''
  arXiv:1606.06687 [hep-th].
 
\bibitem{Jafferis:2015del}
  D.~L.~Jafferis, A.~Lewkowycz, J.~Maldacena and S.~J.~Suh,
  ``Relative entropy equals bulk relative entropy,''
  JHEP {\bf 1606} (2016) 004
  doi:10.1007/JHEP06(2016)004
  [arXiv:1512.06431 [hep-th]].
  
\bibitem{Casini:2013rba}
  H.~Casini, M.~Huerta and J.~A.~Rosabal,
  ``Remarks on entanglement entropy for gauge fields,''
  Phys.\ Rev.\ D {\bf 89} (2014) no.8,  085012
  doi:10.1103/PhysRevD.89.085012
  [arXiv:1312.1183 [hep-th]].

\bibitem{Ghosh:2015iwa}
  S.~Ghosh, R.~M.~Soni and S.~P.~Trivedi,
  ``On The Entanglement Entropy For Gauge Theories,''
  JHEP {\bf 1509} (2015) 069
  doi:10.1007/JHEP09(2015)069
  [arXiv:1501.02593 [hep-th]].

\bibitem{Soni:2015yga}
  R.~M.~Soni and S.~P.~Trivedi,
  ``Aspects of Entanglement Entropy for Gauge Theories,''
  JHEP {\bf 1601} (2016) 136
  doi:10.1007/JHEP01(2016)136
  [arXiv:1510.07455 [hep-th]].

\bibitem{Buividovich:2008gq}
  P.~V.~Buividovich and M.~I.~Polikarpov,
  ``Entanglement entropy in gauge theories and the holographic principle for electric strings,''
  Phys.\ Lett.\ B {\bf 670} (2008) 141
  doi:10.1016/j.physletb.2008.10.032
  [arXiv:0806.3376 [hep-th]].

\bibitem{Donnelly:2011hn}
  W.~Donnelly,
  ``Decomposition of entanglement entropy in lattice gauge theory,''
  Phys.\ Rev.\ D {\bf 85} (2012) 085004
  doi:10.1103/PhysRevD.85.085004
  [arXiv:1109.0036 [hep-th]].

\bibitem{Lin:2018bud}
  J.~Lin and D.~Radicevic,
  ``Comments on Defining Entanglement Entropy,''
  arXiv:1808.05939 [hep-th].
  
\bibitem{Donnelly:2014fua}
  W.~Donnelly and A.~C.~Wall,
  ``Entanglement entropy of electromagnetic edge modes,''
  Phys.\ Rev.\ Lett.\  {\bf 114} (2015) no.11,  111603
  doi:10.1103/PhysRevLett.114.111603
  [arXiv:1412.1895 [hep-th]].

\bibitem{Donnelly:2015hxa}
  W.~Donnelly and A.~C.~Wall,
  ``Geometric entropy and edge modes of the electromagnetic field,''
  Phys.\ Rev.\ D {\bf 94} (2016) no.10,  104053
  doi:10.1103/PhysRevD.94.104053
  [arXiv:1506.05792 [hep-th]].
  
\bibitem{Hollands:2012sf}
  S.~Hollands and R.~M.~Wald,
  ``Stability of Black Holes and Black Branes,''
  Commun.\ Math.\ Phys.\  {\bf 321} (2013) 629
  doi:10.1007/s00220-012-1638-1
  [arXiv:1201.0463 [gr-qc]].
  
\bibitem{Camps:ToAppear}
  J.~Camps, work in progress.

\bibitem{Kirklin:2018gcl}
  J.~Kirklin,
  ``Subregions, Minimal Surfaces, and Entropy in Semiclassical Gravity,''
  arXiv:1805.12145 [hep-th].
  
\bibitem{Engelhardt:2014gca}
  N.~Engelhardt and A.~C.~Wall,
  ``Quantum Extremal Surfaces: Holographic Entanglement Entropy beyond the Classical Regime,''
  JHEP {\bf 1501} (2015) 073
  doi:10.1007/JHEP01(2015)073
  [arXiv:1408.3203 [hep-th]].

\bibitem{Dong:2017xht}
  X.~Dong and A.~Lewkowycz,
  ``Entropy, Extremality, Euclidean Variations, and the Equations of Motion,''
  JHEP {\bf 1801} (2018) 081
  doi:10.1007/JHEP01(2018)081
  [arXiv:1705.08453 [hep-th]].

\bibitem{Sen:2007qy}
  A.~Sen,
  ``Black Hole Entropy Function, Attractors and Precision Counting of Microstates,''
  Gen.\ Rel.\ Grav.\  {\bf 40} (2008) 2249
  doi:10.1007/s10714-008-0626-4
  [arXiv:0708.1270 [hep-th]].

\bibitem{Solodukhin:2011gn}
  S.~N.~Solodukhin,
  ``Entanglement entropy of black holes,''
  Living Rev.\ Rel.\  {\bf 14} (2011) 8
  doi:10.12942/lrr-2011-8
  [arXiv:1104.3712 [hep-th]].

\bibitem{Sen:2012dw}
  A.~Sen,
  ``Logarithmic Corrections to Schwarzschild and Other Non-extremal Black Hole Entropy in Different Dimensions,''
  JHEP {\bf 1304} (2013) 156
  doi:10.1007/JHEP04(2013)156
  [arXiv:1205.0971 [hep-th]].

\bibitem{Huang:2014pfa}
  K.~W.~Huang,
  ``Central Charge and Entangled Gauge Fields,''
  Phys.\ Rev.\ D {\bf 92} (2015) no.2,  025010
  doi:10.1103/PhysRevD.92.025010
  [arXiv:1412.2730 [hep-th]].

\bibitem{Casini:2015dsg}
  H.~Casini and M.~Huerta,
  ``Entanglement entropy of a Maxwell field on the sphere,''
  Phys.\ Rev.\ D {\bf 93} (2016) no.10,  105031
  doi:10.1103/PhysRevD.93.105031
  [arXiv:1512.06182 [hep-th]].

\bibitem{Huerta:2018xvl}
  M.~Huerta and L.~A.~Pedraza,
  ``Numerical determination of the entanglement entropy for a Maxwell field in the cylinder,''
  arXiv:1808.01864 [hep-th].

\bibitem{Donnelly:2016mlc}
  W.~Donnelly, B.~Michel and A.~Wall,
  ``Electromagnetic Duality and Entanglement Anomalies,''
  Phys.\ Rev.\ D {\bf 96} (2017) no.4,  045008
  doi:10.1103/PhysRevD.96.045008
  [arXiv:1611.05920 [hep-th]].

\bibitem{Witten:2018lgb}
  E.~Witten,
  ``A Note On Boundary Conditions In Euclidean Gravity,''
  arXiv:1805.11559 [hep-th].
\end{thebibliography}
\end{document}